\documentclass[12pt,preprint]{aastex}
\usepackage{graphicx}
\usepackage{psfig}
\usepackage{natbib}
\usepackage[english]{babel}  
\usepackage{subfigure}
\usepackage{url}

\begin{document}
   \title{Bright hot impacts by erupted fragments falling back on the Sun: UV redshifts in stellar accretion}

\author{F. Reale\altaffilmark{1,2}, S. Orlando\altaffilmark{2}, P. Testa\altaffilmark{3}, E. Landi,\altaffilmark{4}, C. J. Schrijver\altaffilmark{5}  }

 \altaffiltext{1}{Dipartimento di Fisica \& Chimica, Universit\`a di Palermo,
              Piazza del Parlamento 1, 90134 Palermo, Italy
              E-mail: fabio.reale@unipa.it }
\altaffiltext{2}{INAF-Osservatorio Astronomico di Palermo, Piazza del Parlamento 1, 90134 Palermo, Italy}
\altaffiltext{3}{Harvard-Smithsonian Center for Astrophysics, 60 Garden Street, MS 58, Cambridge, MA 02138, USA}
\altaffiltext{4}{Department of Atmospheric, Oceanic and Space Sciences, University of Michigan, Ann Arbor, MI, 48109, USA}
\altaffiltext{5}{Lockheed Martin Advanced Technology Center, Palo Alto, CA 94304, USA}

            

 
\begin{abstract}  
   {A solar eruption after a flare on 7 Jun 2011 produced EUV-bright impacts of fallbacks far from the eruption site, observed with the Solar Dynamics Observatory. These impacts can be taken as a template for the impact of stellar accretion flows. Broad red-shifted UV lines have been commonly observed in young accreting stars.}
   {Here we study the emission from the impacts in the Atmospheric Imaging Assembly's UV channels and compare the inferred velocity distribution to stellar observations. }
   {We model the impacts with 2D hydrodynamic simulations.}
   {We find that the localised UV 1600\AA\ emission and its timing with respect to the EUV emission can be explained by the impact of a cloud of fragments. The first impacts produce strong initial upflows. The following fragments are hit and shocked by these upflows. The UV emission comes mostly from the shocked front shell of the fragments while they are still falling, and is therefore redshifted when observed from above. The EUV emission instead continues from the hot surface layer that is fed by the impacts.}
   {Fragmented accretion can therefore explain broad redshifted UV lines (e.g. C IV 1550 \AA) to speeds around 400 km/s observed in accreting young stellar objects.}
\end{abstract}


\section{Introduction}

The impact of dense plasma fragments falling back on the solar surface after an eruption has been taken as a template of the impact of accretion flows in young stellar objects \citep{Reale2013a}. In one event on June 7 2011, the impacts occurred far from the eruption site and produced intense EUV brightenings observed by the Atmospheric Imaging Assembly (AIA) on-board the Solar Dynamics Observatory (SDO), thus supporting the scenario where the soft X-ray excess observed in young stars is due to the impact of accretion flows, but also the possibility of fragmented flows, that was overlooked before. 

This event produced also softer brightenings observed in the UV 1600\AA\ and 1700 \AA\ channels of SDO/AIA. 
UV emission from young accreting star systems is a debated issue. Recent work has collected and analysed a large amount of data from T Tauri systems \citep{Ardila2013a,Dupree2014a}. Much attention has been devoted to the C~IV 1550\AA\ line and it has been found that the most common morphology includes a strong, narrow emission component with moderate Doppler shift ($\leq 50$ km/s), either blue or red, together with a weaker, redshifted, broad component, with a tail up to 400 km/s or more, which has been directly associated to the accretion. Whereas UV emission is expected to originate deep at the base of the accretion column, where the ram pressure of the accretion flow equals the local hydrostatic pressure \citep{Drake2005a,Sacco2010a}, 
different explanations have been proposed for this complex line structure, which involve either a complex geometry of the impact region \citep{Romanova2004a} or a structured accretion column \citep{Sacco2010a,Orlando2010a,Ingleby2013a}.
In this work, we use high-speed plasma impacts observed on the Sun as a reference and show that the presence of a steady redshifted C IV line component can be naturally linked to the fragmentation of the accretion flow, and originate from falling plasma that brightens before smashing into the surface.


\section{The observation}
\label{sec:data}

We consider the event analysed in \cite{Reale2013a}. The eruption was initiated by an M2.5 flare from AR 11226 on 7 June 2011, at 6 UT. The details of the data processing and analysis in the EUV channels are described in the Supplementary Materials of \cite{Reale2013a}. Here we focus on the impacts as observed in the UV band, i.e. the 1600\AA\ AIA channel, and compare them to the corresponding emission observed in the EUV 171\AA\ AIA channel.

We summarise the main steps of the data processing and analysis. 
The images of the sequence have exposure times of 2s and 2.9s and cadence of 12s and 24s in the 171\AA\ and 1600\AA\ channel, respectively.
We analyze level 1.5 data. Using the absorption of the impacting fragments in the EUV channels, \citep{Landi2013a} estimated a fragment density between 2 and  $10 \times 10^{10}$ cm$^{-3}$, with a best value of $5 \times 10^{10}$ cm$^{-3}$. 
The trajectories of many infalling fragments are parabolic, as expected for a free fall.
The final fragment speed for many bright impacts has been accurately measured in various independent ways to be in the 300-450 km/s range, with a best value of 400 km/s. The brightenings are well-explained by the dynamics of the fragments and their impact onto the dense chromosphere, and the magnetic field here plays no significant role for the dynamics of several fragments, at variance from other studies \citep{Innes2012a,Carlyle2014a,Dolei2014a,van-Driel-Gesztelyi2014a}.

All of the major impacts have emission signatures in the 1600\AA\ and 1700\AA\ channel. 
\cite{Reale2013a} have analyzed several of the observed impacts, with different sizes and duration. Related brightenings have a duration in the range between 1 and 5 min and a size in the range 5 to 50 arcsec. The smaller and shorter ones are presumably caused by the impact of relatively small and round fragments, with a radius of $\sim 2000-4000$ km. The others by larger elongated fragments or series of them. 

\begin{figure}[htbp]               
 \centering
 \subfigure[]
   {\includegraphics[width=6.cm]{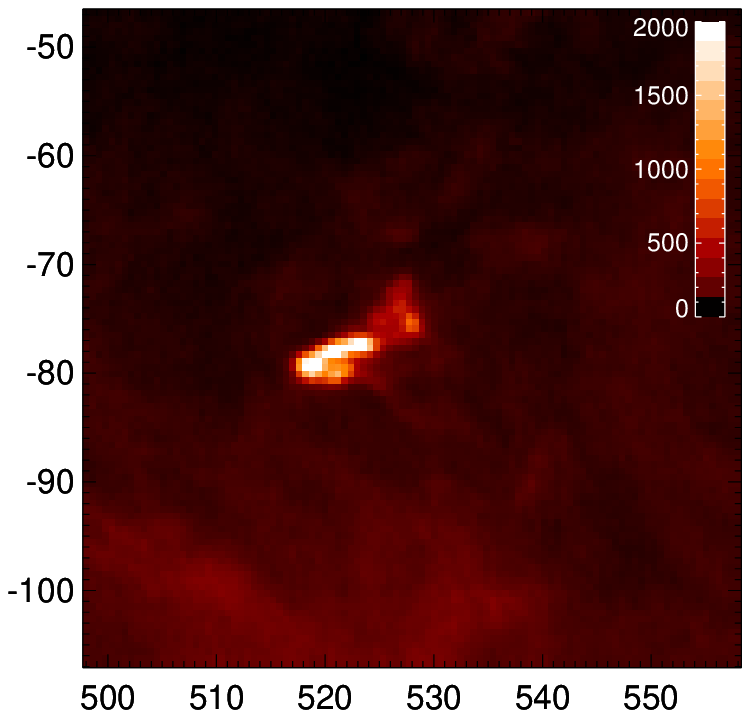}} 
\subfigure[]
   {\includegraphics[width=6.cm]{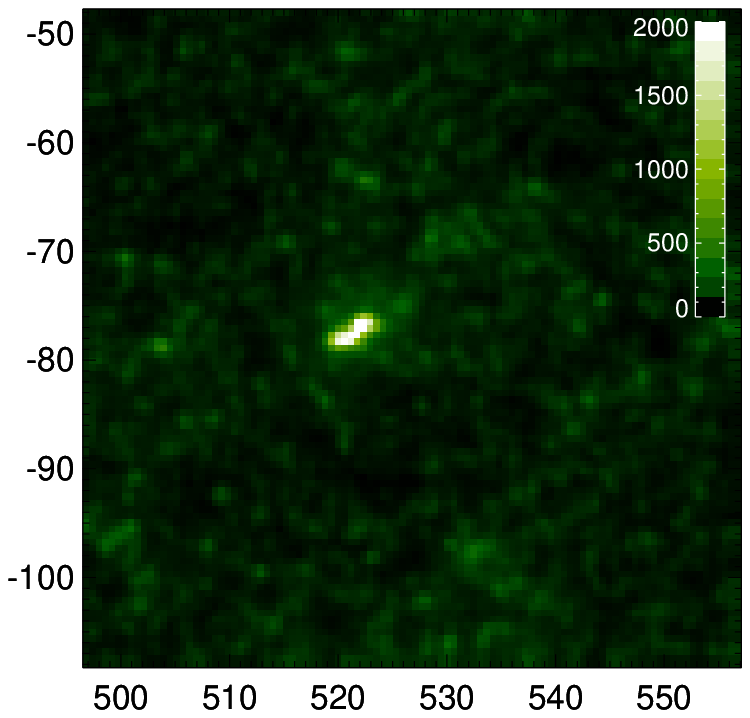}}
\subfigure[]
   {\includegraphics[width=7cm]{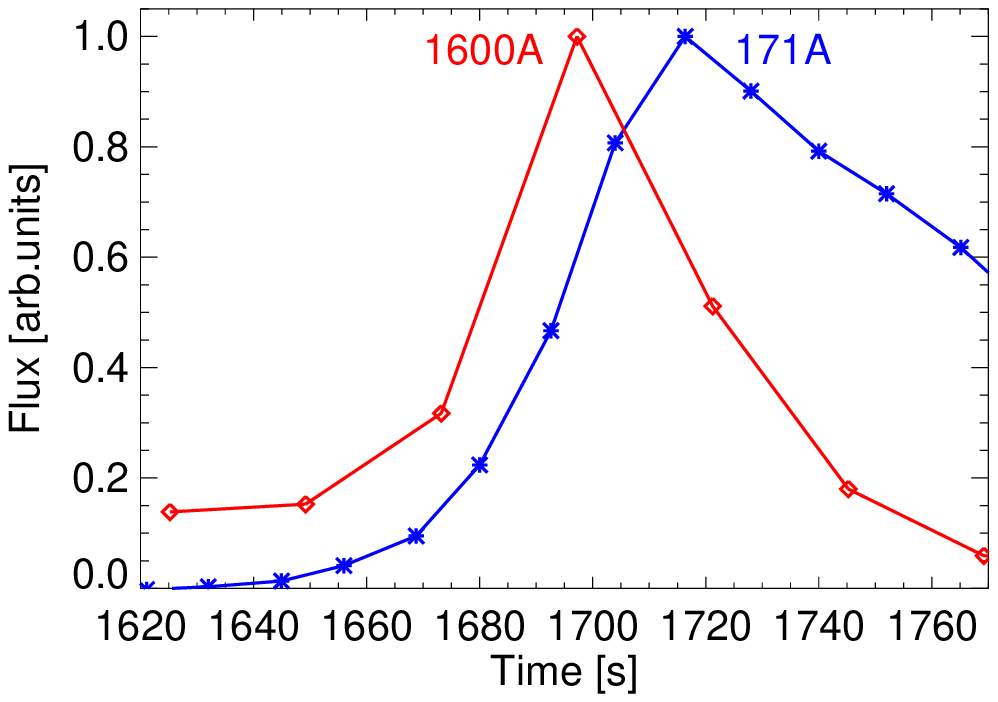}}
   \caption{\footnotesize Impact of a dense fragment falling back on the solar surface: images (a) in the SDO/AIA 171\AA\ channel and (b) in the 1600\AA\ channel at the  peak of the respective light curve. The plot origin is at [X,Y]=[468",-137"] from the disk center. The color palette is data number/pixel/second. (see Movie 1: Time range is between 7:27 and 7:32 UT and the frame cadence is 24 s.) (c) Light curves in the SDO/AIA 171\AA\  (blue) and 1600\AA\ channel (red). Time (in s) starts at 7:00 UT. The light curves are background-subtracted. The (Poisson) data errors are smaller than the symbol size.}
   \label{fig:obs_lc}
\end{figure}

We focus here on one impact that produced a brightening both in the EUV
and in the UV channels, the second shown in \cite{Reale2013a}. It shows very simple morphology and light curves. Many other impacts brighten in both bands \citep[see movies in ][]{Reale2013a}.
Since the falling fragments do not absorb in the 1500-1700\AA\ range, they are not visible in the AIA 1600\AA\ channel before they hit the surface, while their trajectory is visible in absorption in the EUV channels. Typically the 1600\AA\ channel detects  emission in the continuum from the photosphere or lower chromosphere but also a few very intense spectral lines from hotter plasma  (3$\times 10^4$-3$\times 10^5$~K), including the C IV doublet around 1550\AA. 
Fig.~\ref{fig:obs_lc} shows images of the impact brightenings in the 171\AA\ and 1600\AA\ channels (see also Movie~1) and the normalised flux vs time after subtracting an offset value. The offset values (186 and 228 DN/s/pix, respectively) are the minimum values in the time segment. The light curves in the 171\AA\ and 1600\AA\ channels are obtained by integrating the emission in  $36 \times 18$ and $20 \times 10$ pixels area, respectively. The integration area of the 1600\AA\ channel is smaller, because the emission in this channel appears to be more localised than in the other channel. The brightening is more  impulsive in the 1600\AA\ channel with a duration of $\sim 1$ min, while in the 171\AA\ channel it is longer-lasting and has a decay slightly longer than the rise. 
A comparison of the light curves shows that the peak in the 1600\AA\ channel occurs $\sim 30$ s earlier than in the 171\AA\ channel.

\section{The modeling}


To interpret the observation, we simulate dense plasma fragments that
fall through a tenuous corona and their impact onto the dense and
cool chromosphere. As mentioned above, we neglect the role of the
magnetic field. We assume a representative impact
speed of 400 km/s, and a fragment density of $5 \times 10^{10}$
cm$^{-3}$, in the range inferred from observed absorption. The hydrodynamic equations, that include the
gravity stratification and radiative losses from an optically thin
plasma, and the numerical method to solve them are described in
detail in \cite{Reale2013a}. The calculations were performed using
the FLASH code \citep{Fryxell2000a}.

Previous simulations described the simple impact from single spherical and elongated
fragments and of a train of equal spherical fragments \citep{Reale2013a}.
Specifically for this work we have run a new set of simulations
that describe the more general case of the falling of a series of
circular fragments with a random spatial distribution and with a not perfectly vertical trajectory. We use 2D cartesian coordinates in the plane $(x,z)$.
The unperturbed solar atmosphere is hydrostatic
with a hot (maximum temperature $\approx 10^6$ K) and tenuous
($4\times 10^7 < n_{\rm H} < 2\times 10^8$ cm$^{-3}$) corona
\citep{Orlando1996a} linked through a steep transition region to an
isothermal chromosphere, at temperature $10^4$~K.  A perfect balance
of heating and energy losses is assumed in the chromosphere.
Initially, the fragments are in pressure equilibrium with their
surroundings, and their temperature is determined by the pressure
balance across their boundary.


We consider 20 fragments randomly distributed
within a strip $1.5\times 10^9$~cm wide and $10^{10}$~cm long, therefore with an average reciprocal distance of $\sim 9 \times 10^8$ cm. The
fragments hit the chromosphere at an angle $\theta = 75^o$ and
a velocity of 400 km/s. The angle is set only to avoid fragments aligned along the vertical direction and the perfectly symmetric impacts. The fragments have all the same density $n_{fr} = 5 \times 10^{10}$~cm$^{-3}$, but a different radius ranging between $1.4\times 10^8$~cm and $2.6\times 10^8$~cm, i.e. $\leq 1/3$ of their average distance. The end time of the simulation is $ t = 150$~s, that matches approximately the observed duration of the brightenings.

The total size of computational domain is $8\times 10^9$~cm in the $x$
direction and $1.2\times 10^{10}$~cm in the $z$ direction. The grid resolution is very high: an adaptive mesh with 8 nested levels of resolution, that
increases twice at each level with a density-sensitive refinement/derefinement
criterion \citep{Lohner1987a}. The effective resolution is $\approx 1.9\times 10^{6}$~cm at the finest level, corresponding to $\approx 140$ zones per initial width
of the smallest fragment; the effective mesh size is $4096\times 6144$.



From the model results, we synthesize the emission of the optically thin plasma folded
through the instrument standard response in the 171\AA\ and 1600\AA\ SDO/AIA channel, using CHIANTI v.7 \citep{Landi2012a} and assuming coronal abundances \citep{Grevesse1998a}. 
To account for the effect of absorption from optically thick plasma, we set equal to zero the emission from plasma below the transition region in both channels and, only in the 171\AA\ channel, also from plasma denser than $10^{10}~cm^{-3}$. This plasma is found to give a column density of the order of magnitude $\approx 10^{19}$ cm$^{-2}$, the same as that of the downfalling fragments observed in absorption in the EUV channels \citep{Reale2013a}. The fragments instead are not absorbed in the UV channels, due to the very low absorption coefficient at those wavelengths and temperatures.

We also derive the emission profile in the C IV lines with the PINTofALE spectral code \citep{Kashyap2000a}, including the Doppler shifts due to the component of the plasma velocity along the line of sight and excluding the emission from plasma below the transition region.

\begin{figure}[htbp]               
 \centering
 \subfigure[]
   {\includegraphics[width=6.cm]{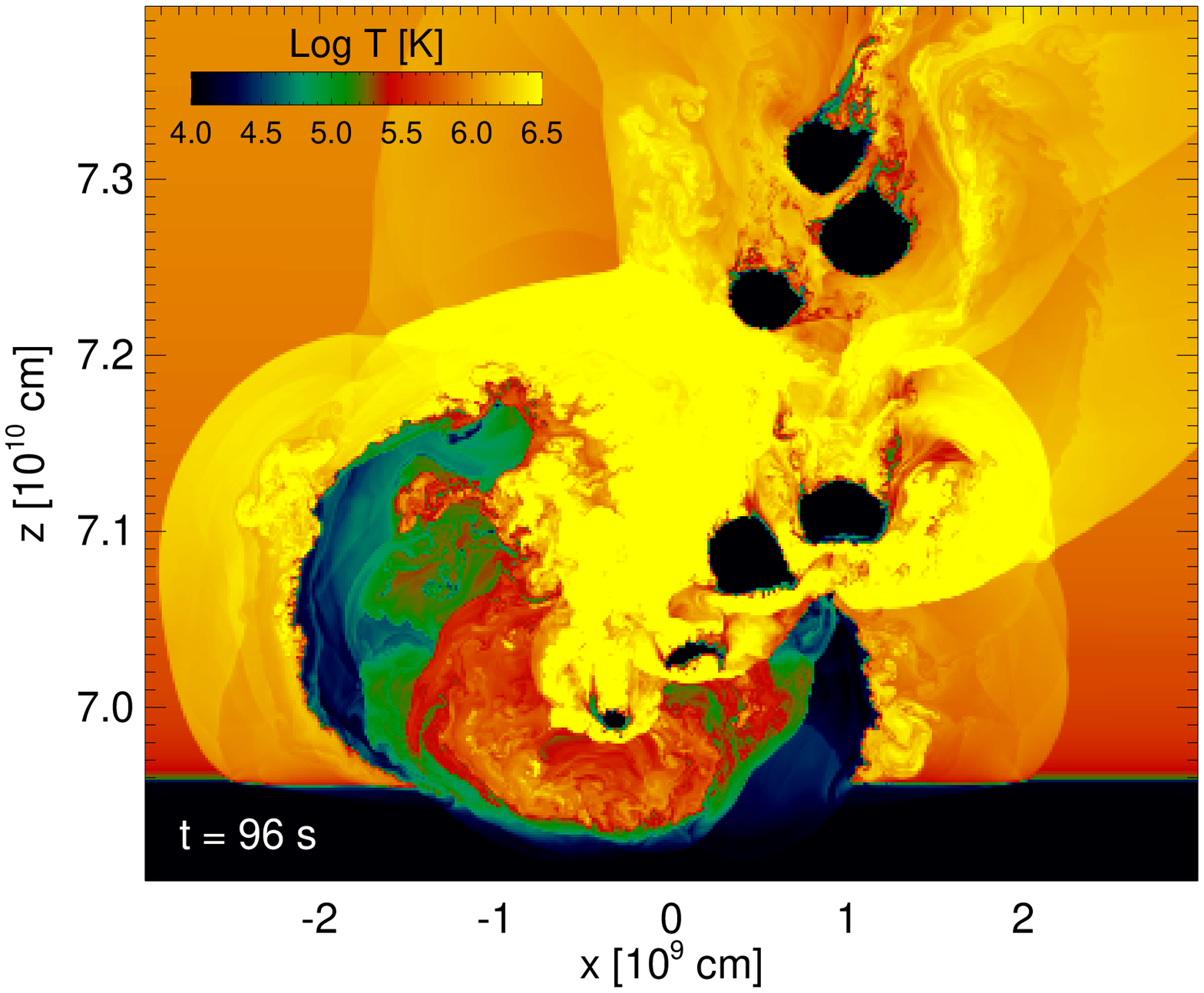}} 
\subfigure[]
   {\includegraphics[width=6.cm]{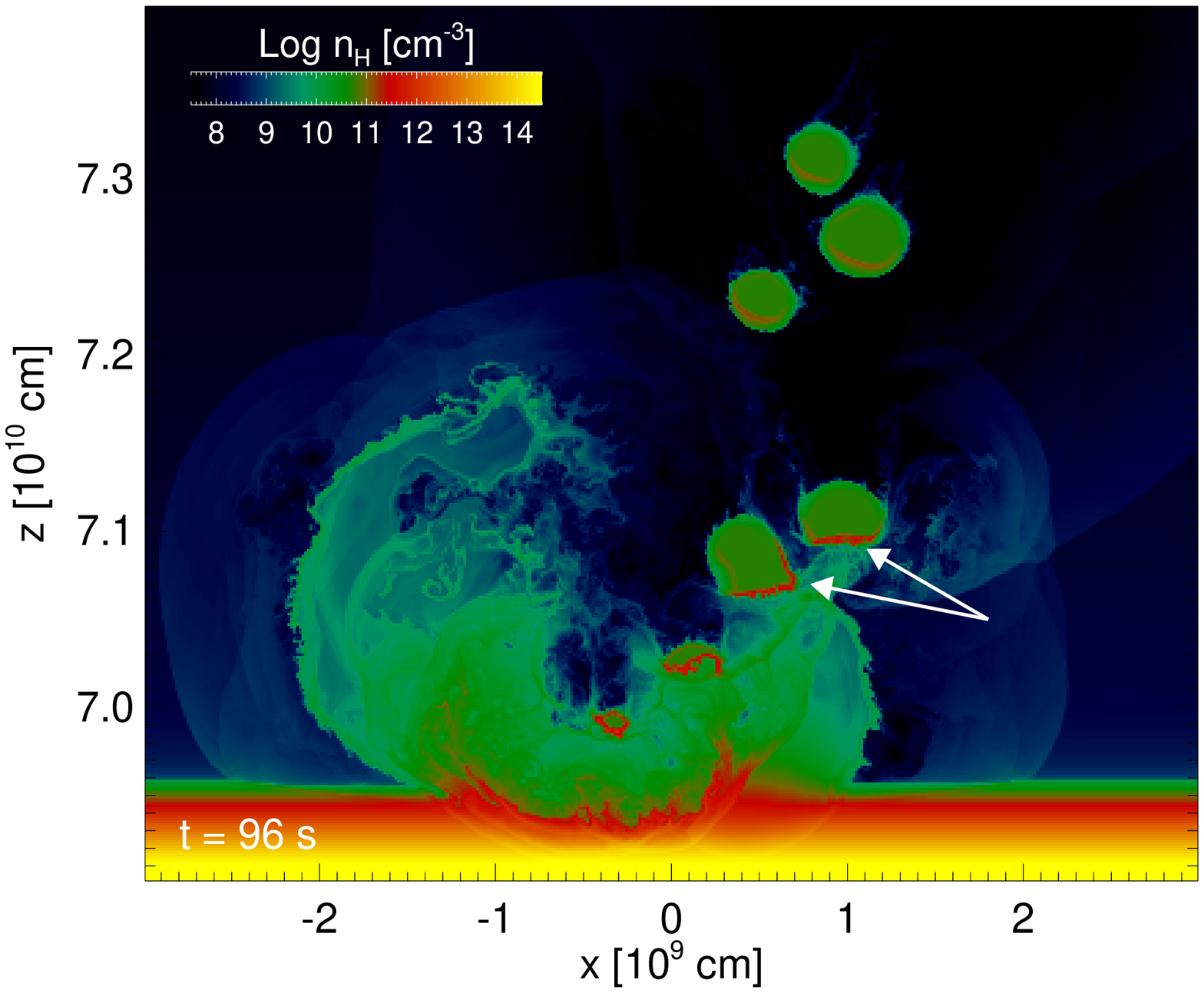}}
\subfigure[]
   {\includegraphics[width=6.cm]{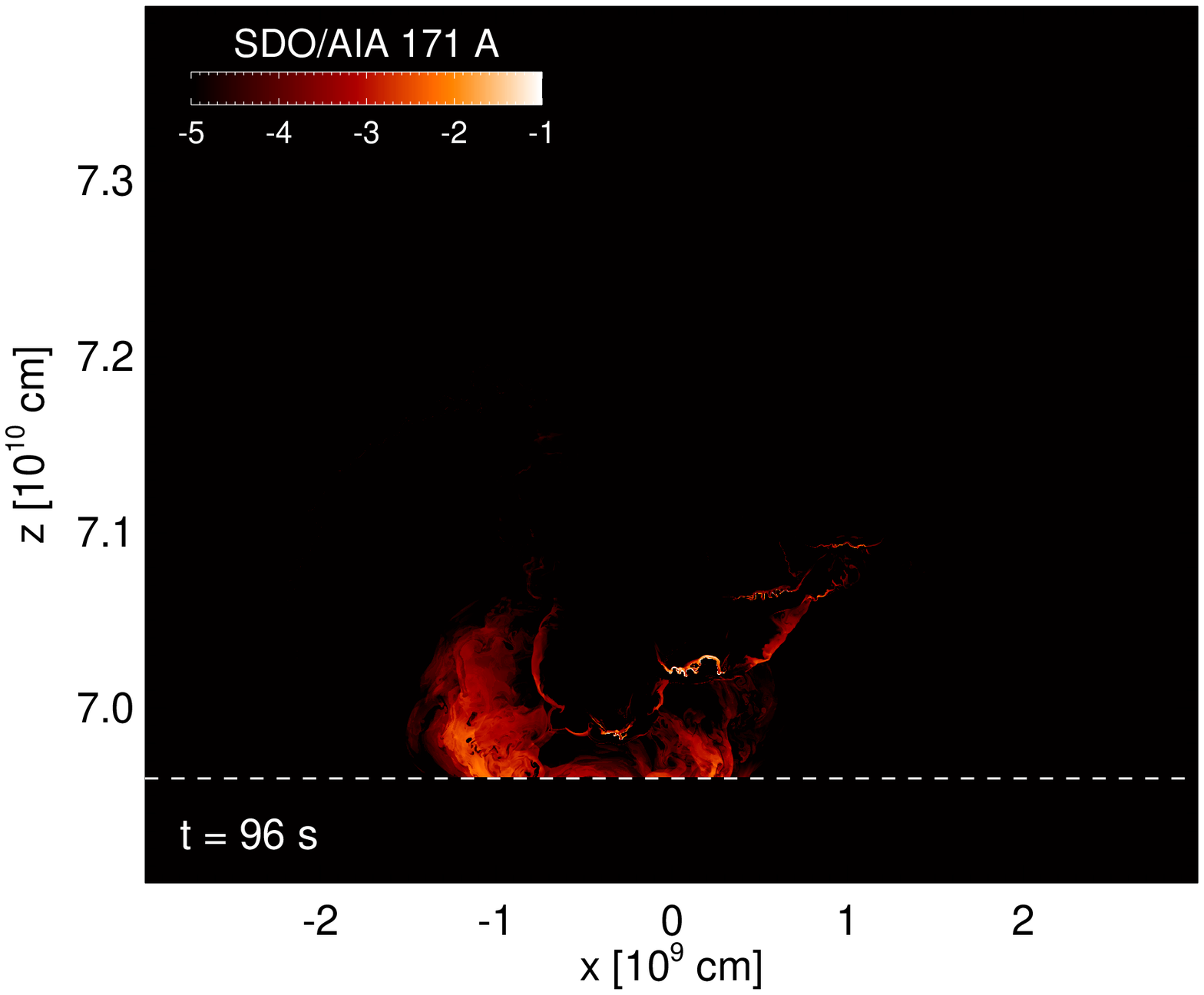}}
   \subfigure[]
   {\includegraphics[width=6.cm]{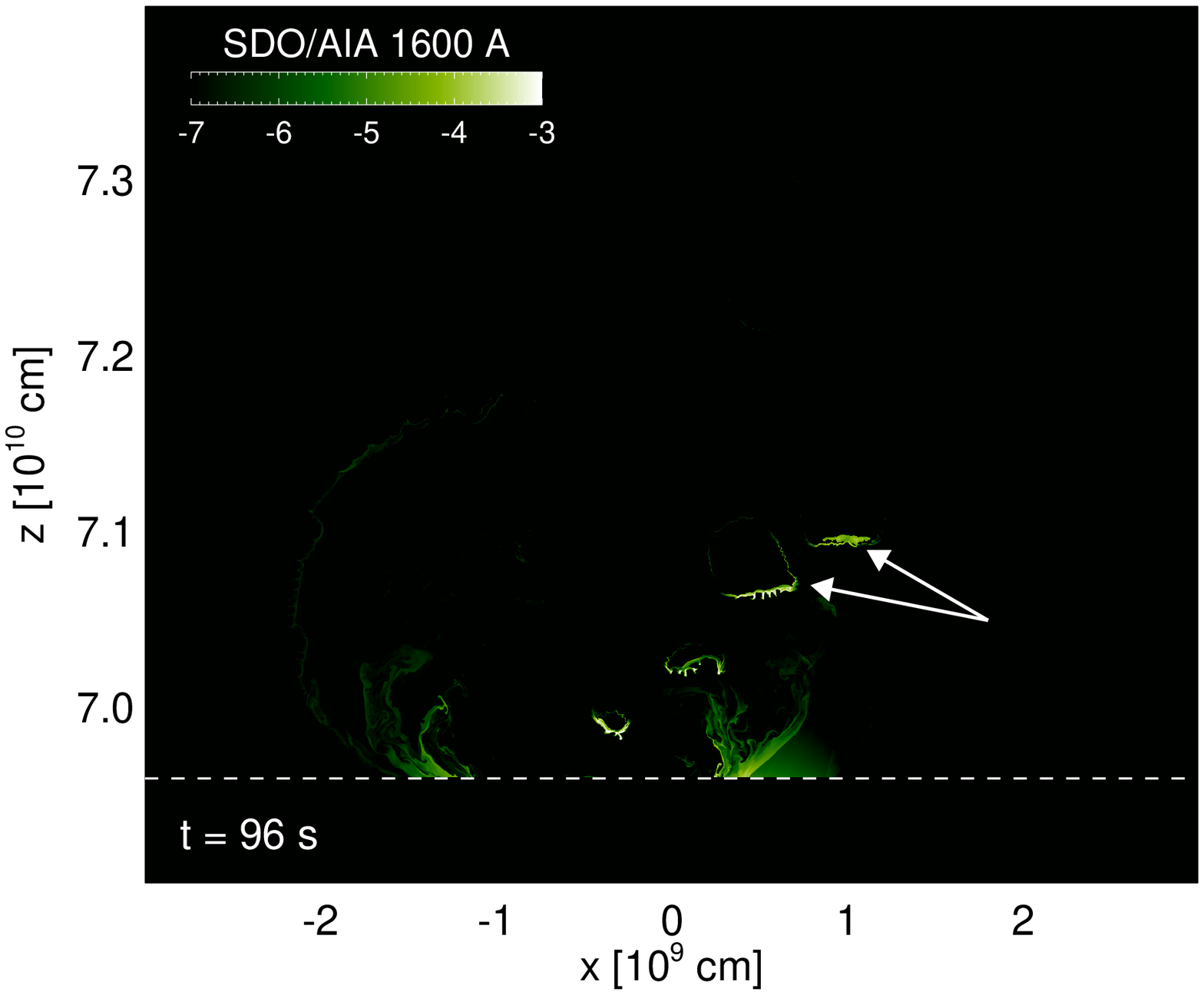}}
   \caption{\footnotesize Maps of the (a) temperature (Movie 2), (b) density (Movie 3) and (c) emission in the SDO/AIA 171\AA\ (Movie 4) and (d) 1600\AA\ (Movie 5) channel from hydrodynamic modeling of randomly distributed falling fragments. The arrows indicate the thin fragment shells which mostly emit in the 1600\AA\ channel. Only the inner domain is shown. The position of the transition region layer is marked in the emission maps (dashed line).
   }
\label{fig:mod_map}
\end{figure}


The impacts of single spherical and elongated fragments and of trains of them have been studied in \cite{Reale2013a}. 
In general, all the modelled fragments free-fall almost undisturbed throughout the tenuous background corona before the impact. Then they smash against the chromospheric layers where the pressure equals the ram pressure of the infalling fragment with a local density $\sim 10^{12}$ cm$^{-3}$. 
The overpressure drives a huge expanding surge backwards in the corona, that is heated to above 1 MK but cools down rapidly to $T < 0.1$ MK in less than 30 s. 

The emission of an impact in the 171\AA\ channel begins deep in the chromosphere and is not observable from outside, and later it extends above the transition region when it bounces back. It remains bright for a very short time ($\sim 20$ s); then it fades out rapidly, due to the cooling and rarefaction. Only a small core of the surge near the surface is bright. 
Most of the emission comes from original fragment material originally. The impact of an elongated fragment lasts longer and drives a larger outflow to a distance of $\sim 10000$ km, bright in the EUV for about 1 min.
Further fragments in column stop the ejection of plasma above the impact but feed considerably the ejection at the sides of the impact point. The bright spot becomes therefore larger and lasts longer because of the longer feeding.

The simulation with randomly distributed falling fragments shows similar general features as the previous more simplified cases (see Movies 2 and 3). The small deviation from the vertical trajectory determines an asymmetric evolution of the impacts. The surge that bounces back is more prominent on the other side with respect to the vertical axis. The random impacts make the evolution complex, with several interacting but very faint bow shocks propagating upwards. 
After the first impact we long see both downflows and upflows at the same time. Fig.~\ref{fig:mod_map} shows maps of temperature and density at time t = 98 s (from the beginning of the simulation) in the inner domain. After the first impacts, hot ($\sim 10^6$ K) and dense ($\sim 10^{10}$ cm$^{-3}$) plasma is found in the surge only close to the impact region, while the outer expanding shells are still dense but cooler ($\log T \sim 5.5$). After later impacts, a uniformly dense layer accumulates just above the chromosphere. Part of this layer is still heated to about $10^6$ K by the late impacts.
Special attention has to be devoted to the fragments following the first ones. Before they impact, while they are still approaching the surface, they are hit by the surging upflows. Their front shell is shocked by the collision and compressed to very high density ($\log n \sim 12$) and high temperature ($\log T \sim 5.5$). 

Fig.~\ref{fig:mod_map} and Movie 4 show that, during the impacts, the emission in the EUV channel (171\AA) above the transition region comes mostly from the upflowing surge, as in the previous simulations \citep{Reale2013a}.
Also a small part of the UV 1600\AA\ emission comes from the surge, but most of the emission has a different origin (Fig.~\ref{fig:mod_map} and Movie 5): the thin shocked front shell of the falling fragments before they hit surface (arrows in Fig.~\ref{fig:mod_map}).

\begin{figure}[htbp]               
 \centering
   {\includegraphics[width=8cm]{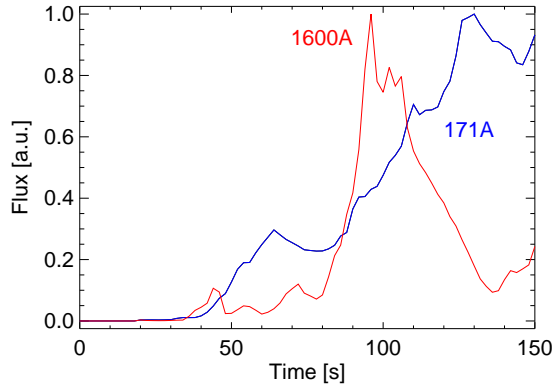}}
   \caption{\footnotesize Light curves from a hydrodynamic simulation of randomly distributed falling fragments in the SDO/AIA 171\AA\ channel (blue) and 1600\AA\ channel (red) (see Fig.~\ref{fig:mod_map}), to be compared with the observed one (Fig.\ref{fig:obs_lc}). }
\label{fig:mod_lc}
\end{figure}

This qualitative difference lead to different light curves.
We extract the light curves in the EUV 171\AA\ and the UV 1600~\AA\ channels from the simulations as we do from the data, by integrating  the emission in the impact region above the chromosphere (Fig.~\ref{fig:mod_lc}), to be compared with the observed ones (Fig.~\ref{fig:obs_lc}). In Fig.~\ref{fig:mod_lc} the peak in the 1600\AA\ channel anticipates the one in the 171\AA\ channel by $\sim 30$ s, in agreement with the observation. The reason is that most of the 1600\AA\ emission comes from falling fragments shocked by the strong initial surge; as the impacts continue, the fragments are hit with less strength, and their emission decreases. The impacts instead continue to feed the hot and dense layer that forms just above the transition region and causes the 171\AA\ emission to increase later. Also, the ratio of the peak intensity is in agreement with the observation (the 171\AA\ peak is a factor 3 higher than the 1600\AA\ peak).

\begin{figure}[htbp]               
 \centering
 \subfigure[]
   {\includegraphics[width=8.cm]{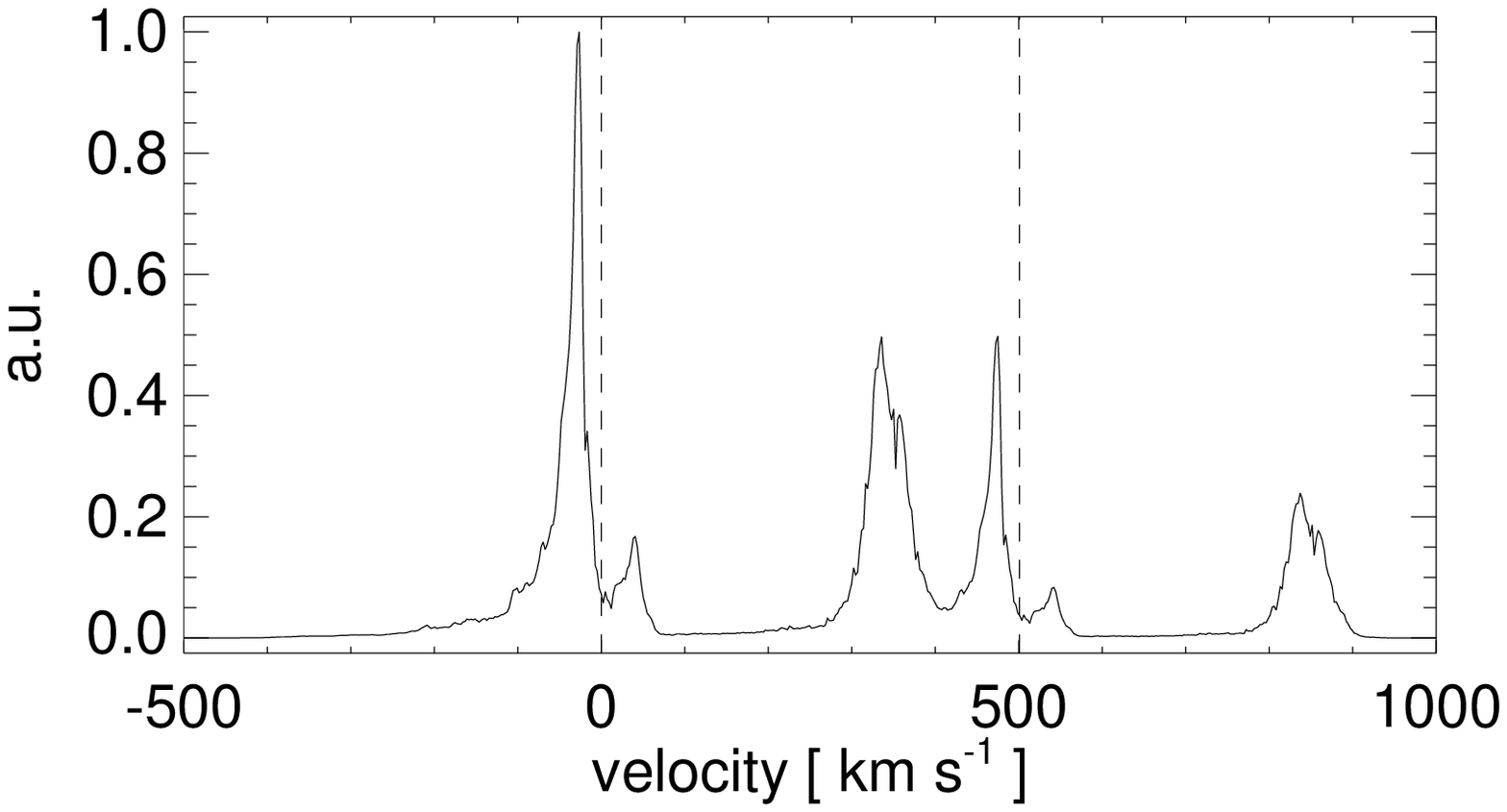}}
\subfigure[]
   {\includegraphics[width=8.cm]{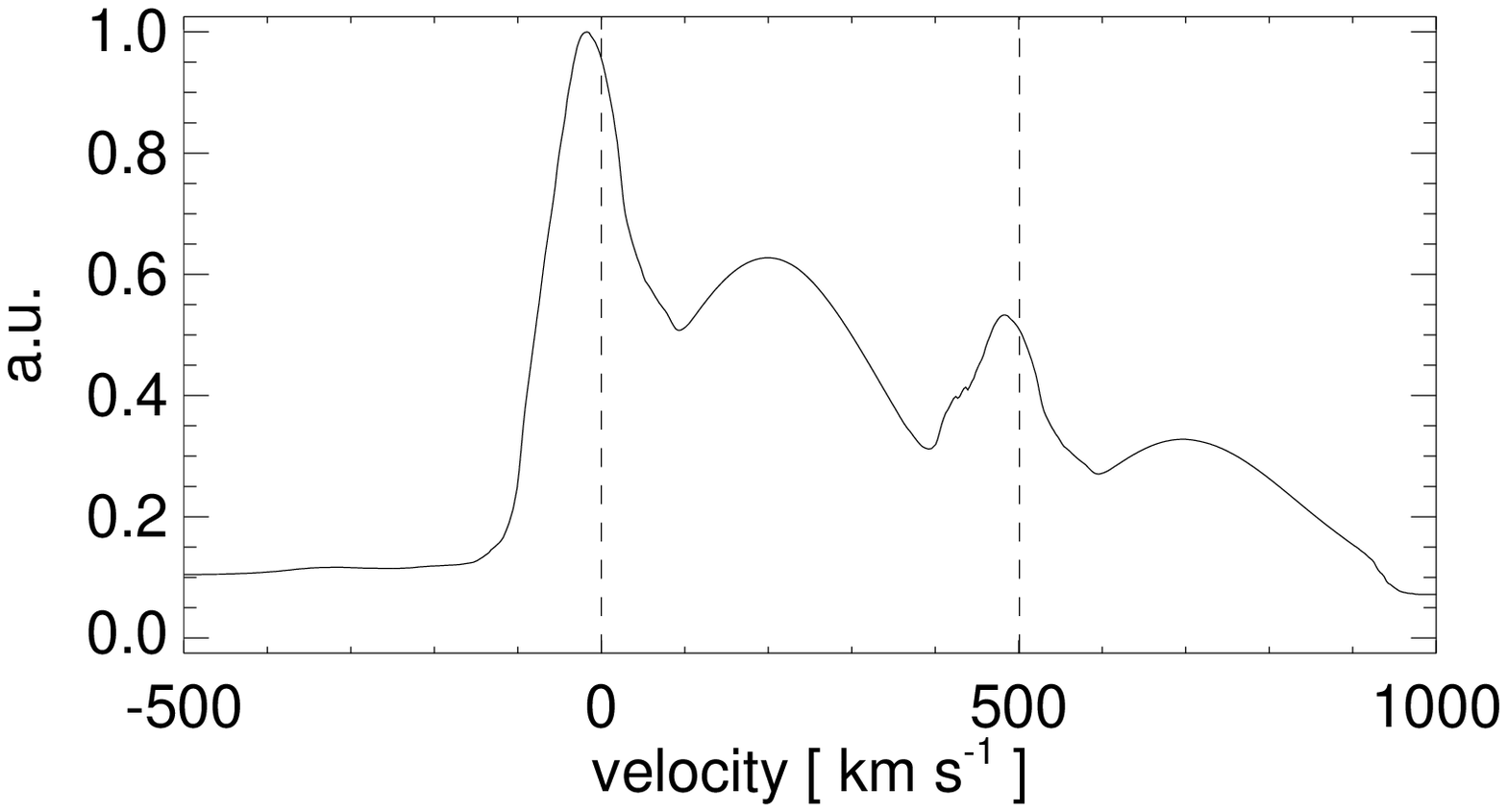}}
   \caption{\footnotesize (a) Synthetic spectrum of the C IV lines from hydrodynamic modeling of randomly distributed falling fragments. (b) Same spectrum folded with Gaussians to account for different velocity angles to the line of sight (see text).}
\label{fig:mod_spec}
\end{figure}

We now investigate more the UV emission in the C IV line. Fig.~\ref{fig:mod_spec}(left panel) shows the spectrum of the C IV 1550 \AA\ doublet integrated over the whole domain and time of the simulation. The line width includes the Doppler shifting as observed from above the stellar surface. The spectrum consists of two identical groups of features, shifted in wavelength, one for each component of the doublet. The main (left) one is also chosen as the velocity reference. Its peak is slightly blue-shifted at a velocity around -50 km/s, and is produced by the material that bounces back (the surge) and therefore moves upwards in the corona after the impact. A smaller peak is located symmetrically at a redshift about $+50$ km/s, but another stronger one is visible at a much higher redshift about $+350$ km/s. This peak is produced by the emission from the front shells of the falling fragments shocked by the up-moving surges. Thus, the redshifts are due to the downfalling material that brightens before it hits surface.


This spectrum corresponds to the case of a single accreting flow observed from a single view angle. In a realistic stellar scenario we might expect a distribution of accreting flows with different view angles, with maximum shift for a flow at the center of the stellar disk and no shift for one at the limb; thus, the larger the velocity, the broader is the distribution. To mimic this more general case, assuming that the observed spectrum is all produced by accretion, we have first shifted the low and high velocity spectrum components derived from our simulation by the average shift ($1/2$) expected for flows uniformly distributed over all possible view angles; then we folded them with bell-shaped (in particular, Gaussian) curves of widths $\sigma = 100$ and 400 km/s, respectively. Each curve describes a possible distribution of orientations of the accreting fragments over the stellar surface with respect to a far observer. Fig.~\ref{fig:mod_spec}(right panel) shows that the line components are vastly broadened, and their profiles are blended. The  redshifted components at a velocity of $\sim 200$ km/s  are very broad with a tail up to about 400 km/s. This pattern is similar to the one observed from the TW Hydra A system \citep{Ardila2013a}.

\section{Conclusions}

We have shown that a simulation of the impacts of several separate fragments falling back on the solar surface is able to explain the brightenings observed on the solar surface in the UV and EUV bands, and also the delay of the EUV 171\AA\ emission from the UV 1600\AA\ emission observed in the light curves of the impacts. So this model shows a good agreement with the observed impacts, including the UV emission. This emission is very important in the framework of the observations of accretion in young stellar systems. Several Doppler-shift patterns were observed in C IV profiles of young stars and our model is able to realistically and quantitatively reproduce at least one of them \citep[TW Hydra A,][]{Ardila2013a}. In our scenario, the presence of a strong highly-redshifted component (400 km/s) is explained by the emission from shells of the dense fragments while they are still falling and are shocked by the strong upflows after the first impacts.  We propose that such redshifted components are a signature of highly fragmented accretion flows, that are heated and compressed by the bounce from previous impacts. The highest speeds to several hundreds km/s mark the speed of the accretion material, and the broadening might be due to a distribution of view angles. Modulations are possible around this basic scenario. This effect might be even magnified in the presence of the expected stronger stellar magnetic fields, where the fragments would be confined and therefore more directly hit by the bouncing fronts. A broadening is expected also if the accreting fragments travel along a twisted magnetic flux tube, because of the related helical motion along the field lines, depending on the view angle. More continuous flows may characterise accretion systems where this pattern is not observed. Optical thickness effects may also contribute to produce different patterns.


\acknowledgements{FR and SO acknowledge support from  italian Ministero dell'Universit\`a e Ricerca. P.T. was supported by contract SP02H1701R from Lockheed-Martin to the Smithsonian Astrophysical Observatory, and by NASA grant NNX10AF29G. The work of EL is supported by several NASA and NSF grants. The software used in this work was in part developed by the DOE-supported ASC/Alliance Center for Astrophysical Thermonuclear Flashes at the University of Chicago. We acknowledge INAF/Osservatorio Astronomico di Palermo for high performance computing resources and support. SDO data supplied courtesy of the SDO/HMI and SOHO/AIA consortia. SDO is the first mission to be launched for NASA's Living With a Star (LWS) Program.}

\bibliographystyle{apj}

\end{document}